\documentclass[amsmath,amssymb,amsfonts,aps,prl,lengthcheck]{revtex4-1}
\usepackage{graphicx}
\usepackage{pdfpages}
\usepackage{lipsum}
\usepackage{float}

\begin{document}
\title{Phase sensing beyond the standard quantum limit with a truncated SU(1,1) interferometer}
\author{Brian E. Anderson$^1$, Prasoon Gupta$^1$, Bonnie L. Schmittberger$^1$, Travis Horrom$^1$, Carla Hermann-Avigliano$^1$, Kevin M. Jones$^2$, and Paul D. Lett$^{1,3}$}
\affiliation{$^1$Joint Quantum Institute, National Institute of Standards and Technology and the University of Maryland, College Park, MD 20742 USA}
\affiliation{$^2$Department of Physics, Williams College, Williamstown, Massachusetts 01267 USA}
\affiliation{$^3$Quantum Measurement Division, National Institute of Standards and Technology, Gaithersburg, MD 20899 USA}
\begin{abstract}
An SU(1,1) interferometer replaces the beamsplitters in a Mach-Zehnder interferometer with nonlinear interactions and offers the potential of achieving high phase sensitivity in applications with low optical powers. We present a novel variation in which the second nonlinear interaction is replaced with balanced homodyne detection. The phase-sensing quantum state is a two-mode squeezed state produced by seeded four-wave-mixing in Rb vapor. Measurements as a function of operating point show that even with $\approx35~\%$ loss this device can surpass the standard quantum limit by 4~dB.
\end{abstract}

\maketitle

Precision phase measurements are invaluable for a wide range of applications, including gravitational wave detection and biological sensing~\cite{10.1038/nphoton.2011.35, PhysRevD.23.1693, 10.1063/1.4724105}. The sensitivity of phase-measuring devices, such as interferometers, is characterized by the uncertainty of a single phase measurement $\Delta\phi$. When using coherent optical fields in a Mach-Zehnder (MZ) interferometer, the phase sensitivity is bounded by the standard quantum limit (SQL) $\Delta\phi=1/\sqrt{N}$, where $N$ is the mean photon number in a single phase measurement.

The phase sensitivity of an interferometer can be improved by using quantum resources~\cite{PhysRevD.23.1693}. By injecting squeezed states~\cite{PhysRevLett.59.278, PhysRevLett.59.2153,Yonezawa1514} or other non-classical states~\cite{PhysRevD.23.1693, PhysRevLett.71.1355, 10.1038/nphoton.2010.268,10.1038/nature02493} into a conventional MZ interferometer, one can achieve phase sensitivities that surpass the SQL.

\begin{figure}
\centering
\includegraphics[scale=0.51]{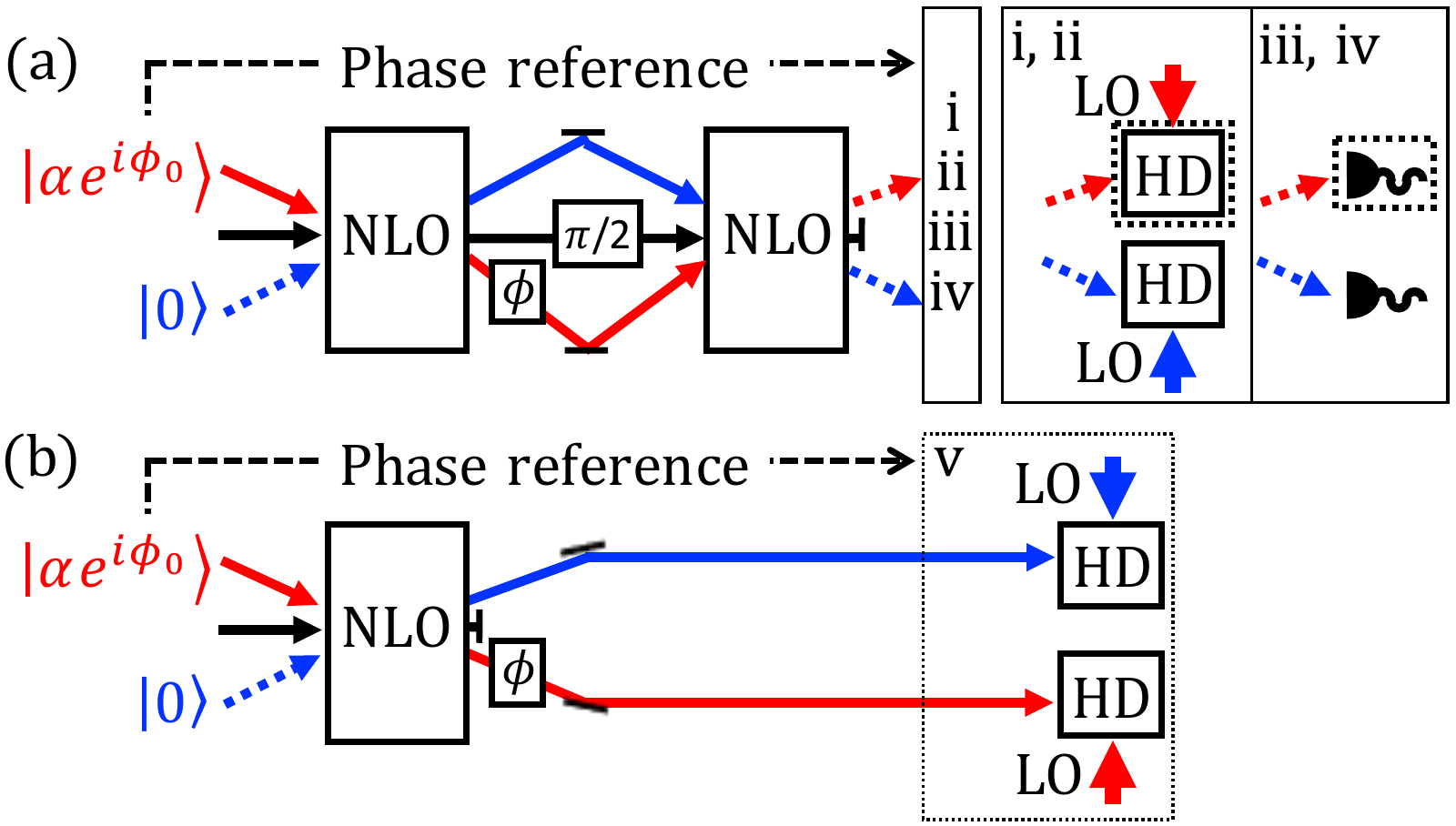}
\caption{Overview of the SU(1,1)-type phase measurement device: A coherent state and a vacuum state are mixed with a pump beam in a nonlinear optical medium (NLO) to produce a two-mode quantum state. A phase shift $\phi$ is applied to the seeded arm. The detection scheme is referenced to the input phase $\phi_0$. (a) The full SU(1,1) interferometer recombines the two-mode squeezed state with a pump beam, phase-shifted by $\pi/2$, in a second nonlinear process and then detects the output modes using (i) two homodyne detectors (HDs), (ii) one HD on the unseeded output mode, (iii) direct intensity detection of only the unseeded mode, or (iv) direct detection in both modes. (b) The truncated SU(1,1) interferometer consists of the first nonlinear interaction, followed by two homodyne detectors (scheme v) that directly collect the two-mode quantum state.}
\label{concept}
\end{figure}

An alternative to injecting quantum states into a MZ interferometer is to replace the passive beamsplitters in such a device with active nonlinear optical elements~\cite{PhysRevA.33.4033}.  A particular configuration that has drawn considerable interest is the SU(1,1) interferometer~\cite{PhysRevA.33.4033}, which is expected to achieve phase sensitivities beyond the SQL even in the presence of loss~\cite{1367-2630-12-8-083014, PhysRevA.86.023844, 10.1038/ncomms4049,PhysRevA.85.023815}. The SU(1,1) interferometer takes two input modes and, using parametric down-conversion or four-wave-mixing (4WM), produces a two-mode squeezed state that is used for phase-sensing, as shown in Fig.~\ref{concept}. We consider the seeded version of the SU(1,1) interferometer, where the phase-sensing two-mode state consists of two bright beams. These beams are then recombined in another nonlinear process, and any changes in the phase sum of the two modes relative to the pump phase can be inferred by detecting the light at the outputs, either by using homodyne detection or direct intensity detection.

We present a variation on the SU(1,1) interferometer that replaces the detection schemes shown in Fig.~\ref{concept}a, including the second nonlinear interaction, with the simpler arrangement shown in Fig.~\ref{concept}b. This scheme eliminates losses associated with the second nonlinear process such as imperfect mode matching of the beams and absorption in the medium. We show theoretically that the phase-sensing ability of this seeded, ``truncated SU(1,1) interferometer'' is the same as that of the full SU(1,1) interferometer shown in Fig.~\ref{concept}a~(i) and surpasses the sensitivity of the direct-detection schemes shown in Fig.~\ref{concept}a~(iii, iv). We build a version of the truncated SU(1,1) interferometer using 4WM and show how the potential phase sensitivity of this device beats the SQL over a range of operating points.

To analyze the phase sensitivity of the full and truncated SU(1,1) interferometers, it is useful to consider the phase-sensing potential of the quantum state separately from the phase sensitivity of the entire device, which depends on the chosen detection scheme~\cite{PhysRevLett.102.040403,1367-2630-16-7-073020,PhysRevA.93.023810}. The phase-sensing potential of the quantum state is identical for both the full and truncated SU(1,1) interferometers and is related to the quantum Fisher information $\mathcal{F}_Q$ according to $\Delta\phi=1/\sqrt{\mathcal{F}_Q}$, where $\mathcal{F}_Q$ is a function of both the two-mode quantum state at the output of the nonlinear medium in Fig.~\ref{concept} and the particular mode(s) in which the phase object(s) are placed~\cite{PhysRevA.85.011801}. The phase sensitivity of the measurement is related to the classical Fisher information $\mathcal{F}_C\le\mathcal{F}_Q$, where $\Delta\phi\geq1/\sqrt{\mathcal{F}_C}$ describes the phase sensitivity of the entire device~\cite{PhysRevA.85.011801}. To optimize the phase-sensing ability of a device, one should choose a measurement scheme such that $\mathcal{F}_C$ is as close as possible to $\mathcal{F}_Q$~\cite{PhysRevA.93.023810}.

Given a measured observable $X$, the phase sensitivity of a measurement device $\Delta\phi$ can be evaluated from the signal-to-noise ratio (SNR),
\begin{equation}
\text{SNR}=\frac{\left[\left(\partial_\phi X\right)\Delta\phi\right]^2}{\Delta^2X},
\label{phasesensigen}
\end{equation}
where $\partial_\phi X$ describes the change in the measurement with respect to a phase shift $\phi$. The minimum detectable phase shift is determined by setting $\text{SNR}=1$. In our experiment, $X$ is the joint quadrature from the balanced homodyne detector signals, given by
\begin{equation}
{X} = {X_p}(\phi_p) + {X_c}(\phi_c),
\end{equation}
where $X_{p,c}$ define the quadratures of the individual field modes of the two-mode squeezed state, and $\phi_p$ and $\phi_c$ are the local oscillator phases at the homodyne detectors, which define the operating point of the interferometer.

For the present purposes, we make the simplifying assumption that the two modes have identical losses (see Ref.~\cite{PhysRevA.78.043816} for a more complete model) and that the seed photon number $|\alpha|^2\gg1$. The sensitivity depends on the operating point set by the local oscillator phases. Specializing to the case where $\phi_c=\pi/2$ (the phase quadrature) but allowing $\phi_p$ to vary, one can show from Eq.~(\ref{phasesensigen}) with $\text{SNR}=1$ that the variance of the phase estimation for the truncated SU(1,1) interferometer is
\begin{equation}
\Delta^2\phi_{\text{tSUI}}=\frac{2\eta+\left(1-2\eta\right)\text{sech}^2(r)-2\eta~\text{sin}(\phi_p)\text{tanh}(r)}{2\eta|\alpha|^2\text{sin}^2(\phi_p)},
\label{phaseTI}
\end{equation}
where $\eta\in[0,1]$ is a loss parameter with $\eta=1$~$(0)$ indicating no loss (total loss), and the squeezing parameter $r=\text{cosh}^{-1}(\sqrt{G})$ depends on the gain $G$ of the nonlinear interaction~\cite{supplementalmat, Briantheory}. For $|\alpha|^2\gg1$, $G$ describes the amplification of the seed, resulting in $G|\alpha|^2$ photons in the seeded mode and $(G-1)|\alpha|^2$ photons in the unseeded mode. We find that, in the ideal lossless case, $\Delta\phi_{\text{tSUI}}$ is equivalent to that of a full SU(1,1) interferometer when using balanced homodyne detection of both outputs, \textit{i.e.}, scheme (i) in Fig.~\ref{concept}a~\cite{supplementalmat, Briantheory}.

\begin{figure}
\centering
\includegraphics[scale=0.41]{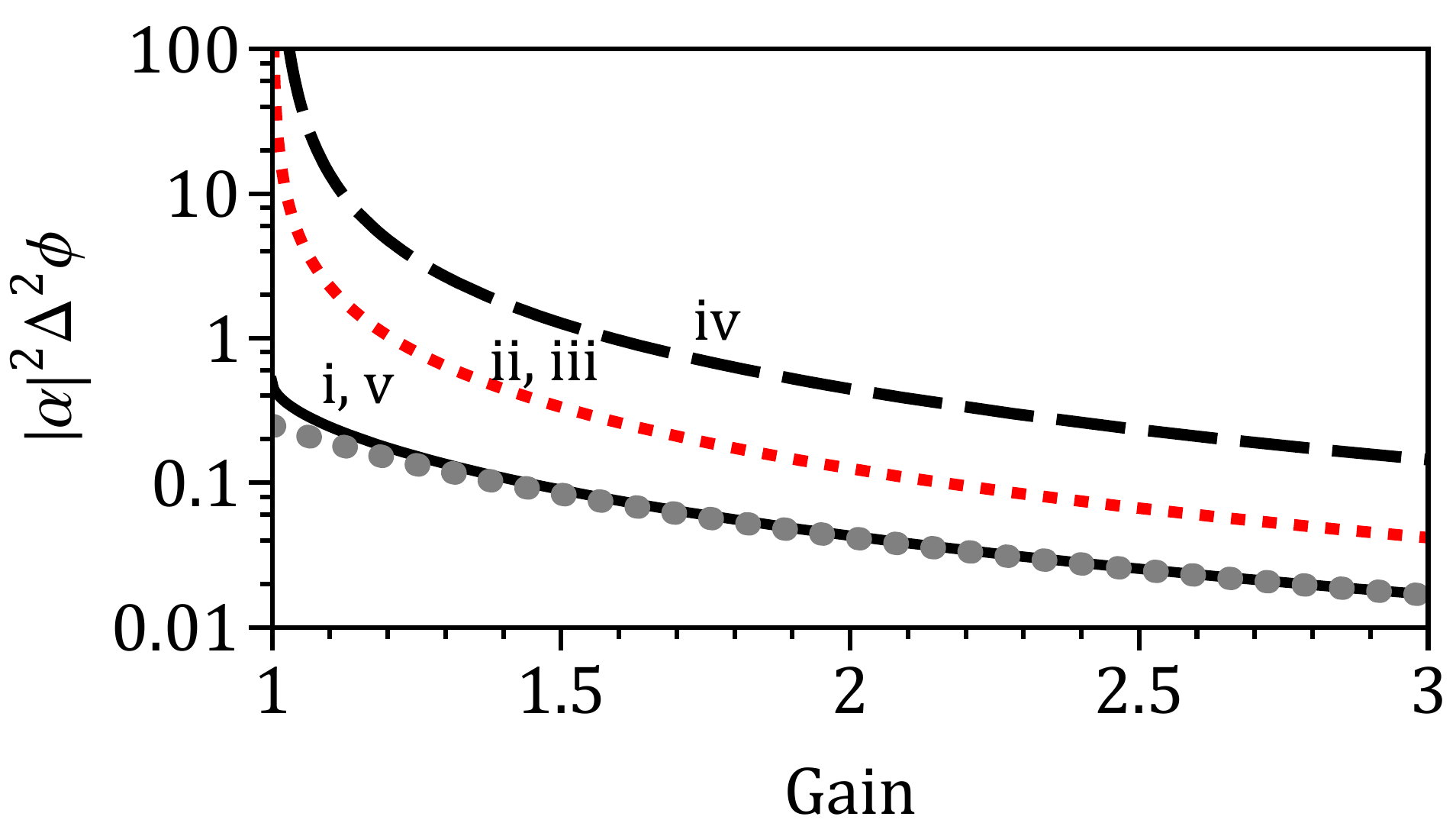}
\caption{The variance of the phase estimation $\Delta^2\phi$ times $|\alpha|^2$, in the case of no loss, as a function of gain, calculated from Eq.~(\ref{phasesensigen}) with $\text{SNR}=1$. The black, solid curve (i, v) defines the optimal phase sensitivities for the full and truncated SU(1,1) interferometers using both HDs (schemes i and v). This curve is also equivalent to $|\alpha|^2/\mathcal{F}_C$ for each detection scheme. The red (short-dashed) and black (long-dashed) curves are the phase sensitivities for the full SU(1,1) interferometer using just a conjugate detector (HD or intensity, schemes ii and iii), and for both intensity detectors (scheme iv), respectively. The gray circles are $|\alpha|^2/\mathcal{F}_Q$ and represent the fundamental phase sensitivity of the two-mode squeezed state.}
\label{sensiplotwithQFI}
\end{figure}

Figure~\ref{sensiplotwithQFI} shows the variance of the phase estimation $\Delta^2\phi$ in the case of no loss as a function of the gain of the nonlinear process that produces the two-mode squeezed state. The variance shown in Fig.~\ref{sensiplotwithQFI} is defined for the best operating point, \textit{e.g.}~curve (v) is for $\phi_p=\pi/2$ in Eq.~\ref{phaseTI}. We also show the variance in phase estimation for the full SU(1,1) interferometer when detecting just the conjugate (unseeded) mode using homodyne detection (scheme ii) or direct intensity detection (scheme iii) as well as direct detection in both modes (scheme iv). (See supplemental materials~\cite{supplementalmat}.) Balanced homodyne detection of the joint quadratures substantially improves the phase sensitivity in the seeded truncated and full SU(1,1) interferometers.

In addition, Fig.~\ref{sensiplotwithQFI} shows the fundamental bound on the phase sensitivity of the two mode quantum state used here, as calculated from the quantum Fisher information $\mathcal{F}_Q$. For the seeded two-mode squeezed state and phase object placement considered here (see Fig.~\ref{concept})~\cite{PhysRevA.93.023810},
\begin{equation}
\mathcal{F}_Q=2\text{cosh}^2\left(r\right)\left[\left(1+2|\alpha|^2\right)\text{cosh}\left(2r\right)-1\right].
\label{FQ}
\end{equation}
We note that if the phase object were placed in the lower-power, vacuum-seeded arm, $\mathcal{F}_Q$ is smaller~\cite{PhysRevA.93.023810}. The result in Eq.~(\ref{FQ}) also requires that the measurement has an external phase reference, as shown in Fig.~\ref{concept}, so that the measured sensitivity is independent of the phase of the input beam~\cite{PhysRevA.85.011801}. One can see that, even at low gain ($G\gtrsim2$), $\Delta^2\phi_{\text{tSUI}}\rightarrow1/\mathcal{F}_Q$. The phase sensing ability of the truncated SU(1,1) interferometer is thus not only equivalent to that of the full SU(1,1) interferometer, but is an optimal measurement choice for sufficiently high gain.

In a practical device the losses play an important role and cause the minimum detectable phase shift to increase. In the full SU(1,1) interferometer the internal losses between the nonlinear interactions have a more detrimental effect on the sensitivity than external losses after the second nonlinear interaction~\cite{PhysRevA.86.023844}. In the truncated SU(1,1) interferometer all losses are effectively internal losses. While the truncated SU(1,1) interferometer avoids additional internal loss arising from coupling the beams into a second nonlinear device, the full SU(1,1) has an advantage of being less sensitive to detector losses~\cite{supplementalmat}.

In the particular case of Gaussian quantum states and homodyne measurements, the classical Fisher information $\mathcal{F}_C$ is related to the measurement observable $X$ by~\cite{PhysRevA.93.023810}
\begin{equation}
\mathcal{F}_C=\frac{\left(\partial_\phi X\right)^2+2\left(\partial_\phi\Delta X\right)^2}{\Delta^2X}.
\label{CFIeq}
\end{equation}
The first term is the inverse of the sensitivity limit determined from the SNR in Eq.~(\ref{phasesensigen}). If the second term is non-zero, one can achieve a better phase sensitivity by using the change in the distribution as a function of phase~\cite{PhysRevLett.99.223602}. In our scheme, the most sensitive operating point corresponds to the minimum of the joint phase quadrature, at which point $\partial_\phi\Delta X=0$.

\begin{figure}
\centering
\includegraphics[scale=0.53]{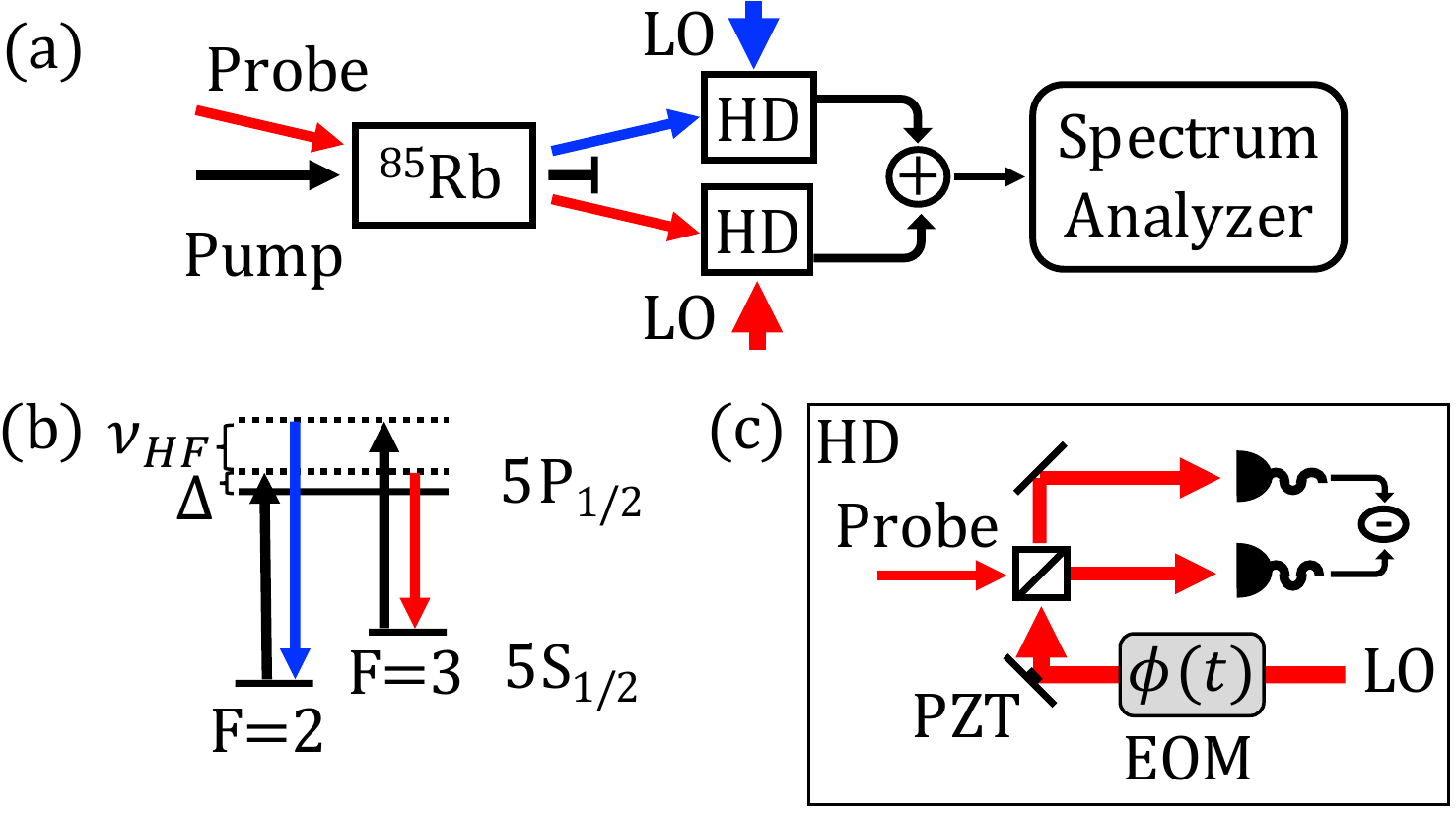}
\caption{(a) The experimental setup consists of a strong pump beam (power ${\approx400}~\text{mW}$, $1/e^2$ beam waist $0.7~\text{mm}$) and orthogonally polarized weak probe beam (power ${\approx100}~\text{nW}$, $1/e^2$ beam waist $0.4~\text{mm}$) propagating at a small angle (${\approx}4~\text{mrad}$) through a warm rubidium vapor (temperature $105~^\circ\text{C}$). After the cell, the pump beam is blocked using a Glan-Taylor polarizer, and the probe and conjugate beams propagate to separate balanced HDs. The local oscillators for each HD are generated using a similar 4WM setup resulting in local oscillator powers of $\approx1~\text{mW}$. The sum of the AC photocurrents is sent to the spectrum analyzer.  (b) We pump the D1 $5S_{1/2}\rightarrow5P_{1/2}$ transition in $^{85}$Rb. Here, $\Delta\approx700~\text{MHz}$ and $\nu_{HF}=3.03~\text{GHz}$ is the hyperfine splitting between the $F=2$ and $F=3$ ground state sublevels. The 4WM generates a conjugate field (blue) that is $6.06~\text{GHz}$ higher in frequency than the applied probe field (red). (c) The probe HD setup consists of a local oscillator (LO) propagating through an electro-optic modulator (EOM) and reflecting off a piezo-electric-transducer-mounted mirror (PZT). The LO mixes with the probe beam on a 50/50 beam splitter, and each output is sent to a subtracting detector. The conjugate HD setup does not contain an EOM.} 
\label{su11setup}
\end{figure}

We construct the truncated SU(1,1) interferometer using 4WM in a $^{85}$Rb vapor cell, as depicted in Fig.~\ref{su11setup}(a, b)~\cite{PhysRevA.78.043816}. The nonlinear interaction among the applied beams and the atoms results in an amplified probe and the creation of a correlated conjugate field. After the cell, the probe and conjugate fields are sent to separate balanced homodyne detectors, as shown in Fig.~\ref{su11setup}c.

To detect a specific quadrature, we use piezo-mounted mirrors in the path of each local oscillator to control the phases of each homodyne setup, $\phi_p$ and $\phi_c$. The AC components ($\approx1~\text{MHz}$) of the two homodyne detectors are summed and sent to a spectrum analyzer. The DC components are used to lock the phase of each homodyne detector.

By locking the homodyne detectors to the DC component of the signal we effectively lock the homodyne detectors to the input phase of the seed beam, indicated by the phase reference line in Fig.~\ref{concept}.  This allows us to assess the potential phase sensing ability of the device without having to build a fully phase stable interferometer.  Likewise, by measuring noise at a convenient high frequency $\Omega=1~\text{MHz}$, we avoid technical noise sources. In the absence of technical noise, the noise floor at $1~\text{MHz}$ would extend to DC. The optimal phase sensing is achieved when both homodyne detectors are locked to the phase quadrature of their input beam. This occurs where $\phi_p$ and $\phi_c$ are locked so that the individual DC homodyne signals are zero and the slopes are the same~\cite{supplementalmat}.

\begin{figure}
\centering
\includegraphics[scale=0.58]{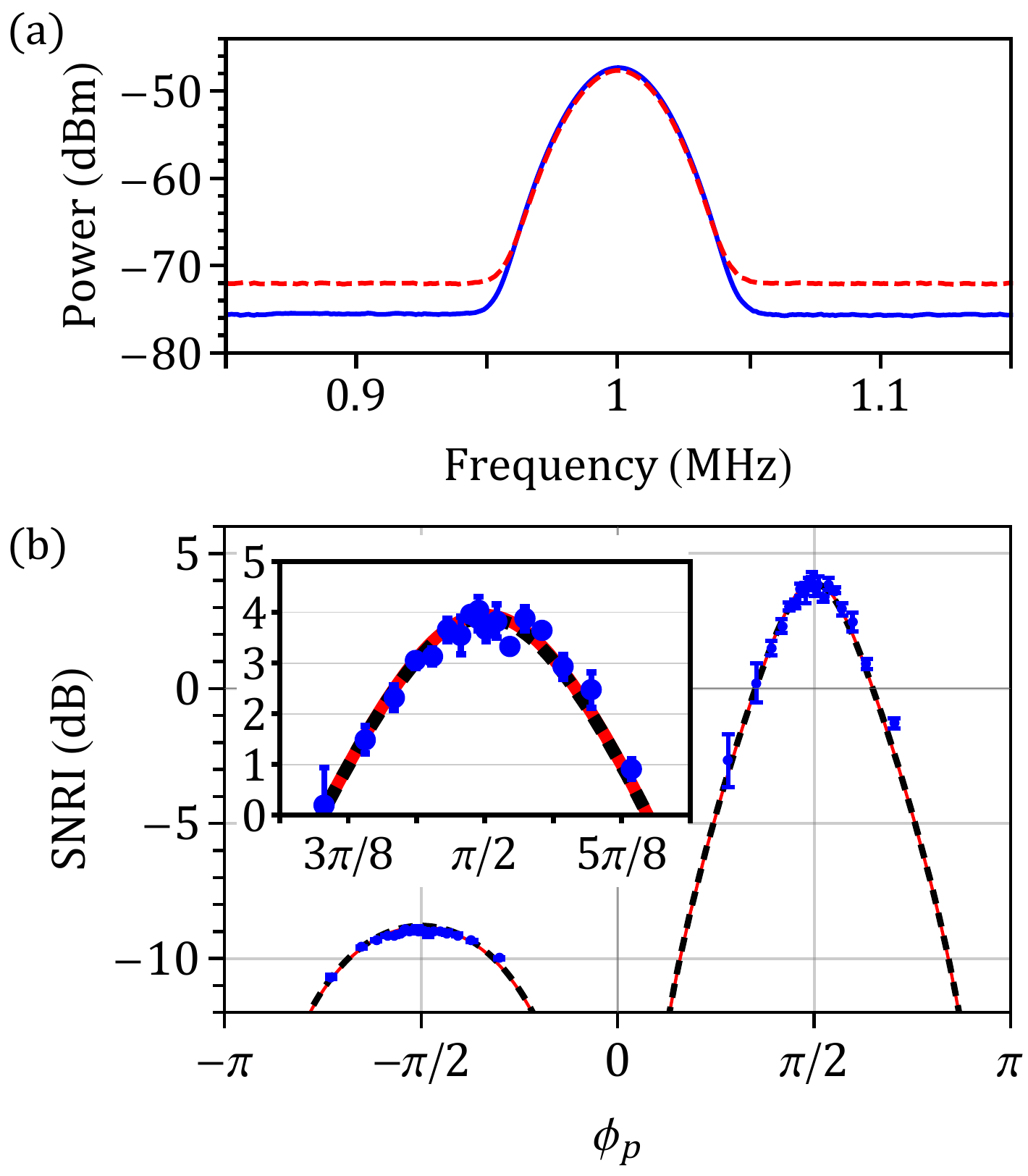}
\caption{(a) An example SNR measurement (resolution bandwidth 30~kHz) using two-mode squeezed beams (blue, solid) and coherent beams (red, dashed), with identical optical power in the phase-sensing (probe) beam path. (b) The SNRI as a function of $\phi_p$. The inset shows the region where we observe an improvement beyond the SQL. The red curve is from Eq.~(\ref{phaseTI}) with best-fit parameters $\eta=0.65$ and intrinsic gain $G=3.3$. The black, dashed curve shows the potential SNRI which could be obtained by additional signal processing as derived from the classical Fisher information in Eq.~(\ref{CFIeq}) using those fit parameters. The uncertainties are statistical, one standard deviation for 10 measurements.}
\label{phasedatafig}
\end{figure}

An electro-optic phase modulator (EOM) imposes a weak sinusoidal phase shift $\phi(t)=\sqrt{2}~\delta\phi~\text{cos}(2\pi\Omega t)$ on the probe local oscillator, where $\delta\phi$ is the RMS amplitude of the phase shift. The signal resulting from this modulation appears on the spectrum analyzer as a peak, shown in Fig.~\ref{phasedatafig}(a), with a signal-to-noise ratio $\text{SNR}_{\text{tSUI}}$. The signal is generated in the probe detection arm and is determined by $\delta\phi$ and the probe and local oscillator powers. The noise level is determined by both detection arms, \textit{i.e.}, by the level of squeezing generated by the 4WM process. Both signal and noise vary with the operating point as set by the local oscillator phases. Absent low-frequency technical noise, the modulation frequency could be lowered while keeping a constant SNR. Thus we use this AC signal to infer a sensitivity to DC phase shifts for comparison to the theory given above.

For comparison to this quantum-enhanced measurement procedure, we follow a similar procedure using coherent fields. We turn off the 4WM process by blocking the pump and increase the intensity of the seed beam to create a coherent state probe beam with the same power as the one generated by the 4WM process, resulting in the dashed curve in Fig.~\ref{phasedatafig}a with signal-to-noise ratio $\text{SNR}_{\text{coh}}$. In this case $\text{SNR}_{\text{tSUI}}-\text{SNR}_\text{coh}\approx4~\text{dB}$.

A previous demonstration of a full SU(1,1) interferometer~\cite{10.1038/ncomms4049} uses the detection scheme defined by curve (ii, iii) in Fig.~\ref{sensiplotwithQFI}, which is not the optimal choice of measurement even in the case of no loss. We note that the conventions used in Ref.~\cite{10.1038/ncomms4049} imply a 3~dB improvement in SNR even at $G=1$, where there are no quantum enhancements. 

To characterize the operation of the truncated SU(1,1) interferometer, we compare the $\text{SNR}_{\text{tSUI}}$ for a particular choice of local oscillator phases to the optimal $\text{SNR}_{\text{coh}}$ (\textit{i.e.}, the SNR when the local oscillator phases are set to give the largest $\text{SNR}_{\text{coh}}$). The phase sensitivities are then related to the SNR improvement ($\text{SNRI}=\text{SNR}_{\text{tSUI}}-\text{SNR}_\text{coh}$) by
\begin{equation}
\text{SNRI}=-10~\text{log}_{10}\frac{\Delta^2\phi_{\text{tSUI}}}{\Delta^2\phi_{\text{coh}}},
\label{SNRI}
\end{equation}
where $\Delta^2\phi_{\text{coh}}=1/(2\eta G|\alpha|^2)$ is the optimal phase sensitivity achievable using coherent beams.

To measure the SNRI as a function of operating point, we lock $\phi_c$ to detect the phase quadrature of the conjugate beam and scan $\phi_p$. The resulting variation in SNRI is shown in Fig.~\ref{phasedatafig}b. The peaks at $\phi_p=\pm\pi/2$ correspond to the operating points with the maximum signal. At $\phi_p=\pi/2$ the joint quadrature noise is at its minimum (maximum squeezing). At $\phi_p=-\pi/2$ the joint quadrature noise is at its maximum (maximum anti-squeezing). The solid curve is a theoretical fit of Eq.~(\ref{phaseTI}) to the data with best-fit parameters $\eta=0.65$ and intrinsic gain $G=3.3$. The experimentally measured gain is $2.7$, which includes some loss in the vapor cell. For simplicity, we take both arms to have identical loss parameters $\eta$.

The classical Fisher information for this detection scheme in the presence of loss is shown by the dashed line in Fig.~\ref{phasedatafig}b, which is calculated from Eq.~(\ref{CFIeq}) using the fit parameters defined above. The close overlap of this curve with that derived from $\Delta^2\phi_{\text{tSUI}}$ implies that the second term in Eq.~\ref{CFIeq} is negligible compared to the first.

The standard quantum limit, $\Delta\phi_{\text{SQL}}$, is usually taken to mean the best sensitivity of an MZ interferometer with intensity detection at both output ports. We compare the truncated SU(1,1) interferometer to a truncated version of a MZ in which the second beamsplitter is replaced with homodyne detectors. The measured signal $X$ is taken to be the difference in the quadrature signals from the two homodyne detectors. At its optimum operating point this configuration has the same sensitivity as the standard MZ. The SQL depends on the mean number of photons used in the measurement. In the seeded SU(1,1) the two beams have different photon numbers, $N_p$ and $N_c$ (with $N_c<N_p$). Taking the number of photons going through the phase object as the essential resource, we select as the SQL the sensitivity of an ideal MZ with $N_p=\eta G|\alpha|^2$ photons detected in the phase-sensing arm, \textit{i.e.} $\Delta^2\phi_{\text{SQL}}=1/(2N_p)$.

To compare our results to the SQL, we need to compare the measured $\text{SNR}_{\text{coh}}=(\delta\phi)^2N_p$ to that expected for the truncated MZ. We determine $N_p$ in terms of the probe power and estimated losses, and we find that the measured $\text{SNR}_{\text{coh}}$ agrees with the expected value~\cite{supplementalmat}. Thus we can conclude that the SNRI measured here shows that the truncated SU(1,1) achieves a 4~dB improvement in $\Delta^2\phi$ over the SQL including losses. Further, if we could eliminate all detection losses for the coherent state measurement~\cite{supplementalmat}, then $\Delta^2\phi_{\text{tSUI}}$, which includes losses, surpasses that by $3~\text{dB}$.

We have constructed a novel variation on the SU(1,1) interferometer that removes one nonlinear interaction, and we have measured the SNR of this device as a function of operating point. This arrangement provides a simpler means to achieve quantum-enhanced phase sensitivities, and even with $\approx35~\%$ loss we have demonstrated up to 4~dB improvement over the comparable classical device.

We gratefully acknowledge the support of the National Science Foundation and the Air Force Office of Scientific Research, as well as discussions with P. Barberis-Blostein, D. Fahey, and E. Goldschmidt.

\bibliography{SU11refs}

\newpage
\widetext
\clearpage
\setcounter{equation}{0}
\setcounter{figure}{0}
\setcounter{table}{0}
\setcounter{page}{1}
\makeatletter
\renewcommand{\theequation}{S\arabic{equation}}
\renewcommand{\thefigure}{S\arabic{figure}}
\renewcommand{\bibnumfmt}[1]{[S#1]}
\renewcommand{\citenumfont}[1]{S#1}
\begin{center}
\textbf{Phase sensing beyond the standard quantum limit with a truncated SU(1,1) interferometer: Supplemental Material}\\
\vspace{4mm}
{Brian E. Anderson$^1$, Prasoon Gupta$^1$, Bonnie L. Schmittberger$^1$, Travis Horrom$^1$, Carla Hermann-Avigliano$^1$, Kevin M. Jones$^2$, and Paul D. Lett$^{1,3}$}\\
\vspace{4mm}
\textit{$^1$Joint Quantum Institute, National Institute of Standards and Technology and the University of Maryland, College Park, MD 20742 USA}\\
\textit{$^2$Department of Physics, Williams College, Williamstown, Massachusetts 01267 USA}\\
\textit{$^3$Quantum Measurement Division, National Institute of Standards and Technology, Gaithersburg, MD 20899 USA}
\end{center}

\subsection{Deriving the phase sensitivity for the truncated SU(1,1) interferometer}
\begin{figure}
\centering
\includegraphics[scale=0.74]{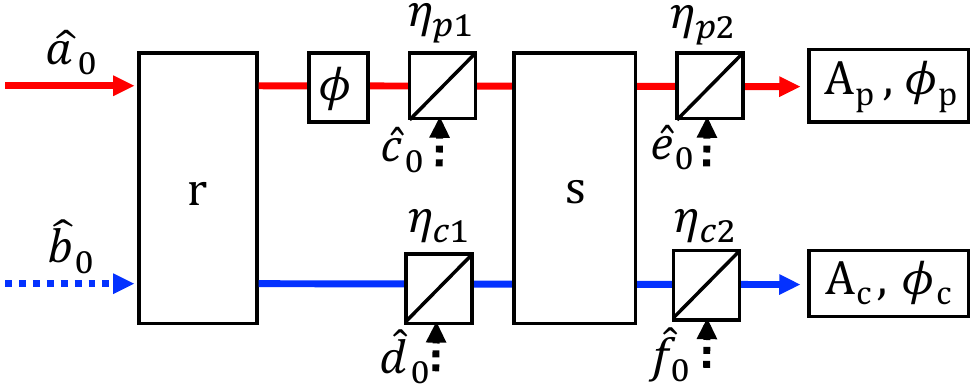}
\caption{A schematic of the model used to derive the phase sensitivity. A coherent state and a vacuum state undergo a squeezing operation of squeezing parameter $r$. The probe (upper) arm undergoes a phase shift $\phi$. Loss is modeled using beamsplitters of reflectivity $1-\eta_{j}$. Homodyne detection is characterized by a classical amplification $A_j$ and a phase $\phi_j$.}
\label{modelsetup}
\end{figure}

We consider two input modes $a$ and $b$, as shown in Fig.~\ref{modelsetup}. Mode $a$ is seeded with a coherent state, where $\left<\hat{a}_0^\dagger\hat{a}_0\right>=|\alpha|^2$ is the seed photon number, and mode $b$ is seeded with a vacuum state~$\left[\text{S1}\right]$. The input modes, described by $\vec{v}$, undergo a squeezing operation $\hat{U}$, as shown in Fig.~\ref{modelsetup}, where for a squeezing parameter $r$,
\begin{equation}
\hat{U}\vec{v}=
\begin{pmatrix}
\text{cosh}(r) & 0 & 0 & \text{sinh}(r)\\
0 & \text{cosh}(r) & \text{sinh}(r) & 0\\
0 & \text{sinh}(r) & \text{cosh}(r) & 0\\
\text{sinh}(r) & 0 & 0 & \text{cosh}(r)\\
\end{pmatrix}
\begin{pmatrix}
\hat{a}_0\\
\hat{a}_0^\dagger\\
\hat{b}_0\\
\hat{b}_0^\dagger\\
\end{pmatrix}.
\end{equation}
The probe (seeded) arm undergoes a phase shift $e^{i\phi}$. We treat loss in terms of beam splitters of transmission $\eta_j$, as shown in Fig.~\ref{modelsetup}, where each beam splitter also injects vacuum noise, described by the mode operators $\hat{c}_0$, $\hat{d}_0$, $\hat{e}_0$, and $\hat{f}_0$. For the full SU(1,1) interferometer, the beams undergo a second squeezing operation of squeezing parameter $s$, which we typically take to be $s=-r$. For the truncated SU(1,1) interferometer, $s=0$. At the output of the second squeezer, the modes of the upper (probe) and lower (conjugate) arms are described by the operators $\hat{a}_f$ and $\hat{b}_f$, where
\begin{multline}
\hat{a}_f=i\hat{e}_0\sqrt{1-\eta_{p2}}+\sqrt{\eta_{p2}}\Bigg\{\text{cosh}(s)\left[i\hat{c}_0\sqrt{1-\eta_{p1}}+\sqrt{\eta_{p1}}\left(\hat{a}_0e^{i\phi}\text{cosh}(r)+e^{i\phi}\text{sinh}(r)\hat{b}_0^\dagger\right)\right]+\\
\text{sinh}(s)\left[\sqrt{\eta_{c1}}\left(\hat{a}_0\text{sinh}(r)+\text{cosh}(r)\hat{b}_0^\dagger\right)-i\sqrt{1-\eta_{c1}}\hat{d}_0^\dagger\right]\Bigg\}
\label{af}
\end{multline}
and
\begin{multline}
\hat{b}_f=i\hat{f}_0\sqrt{1-\eta_{c2}}+\sqrt{\eta_{c2}}\Bigg\{\text{cosh}(s)\left[i\hat{d}_0\sqrt{1-\eta_{c1}}+\sqrt{\eta_{c1}}\left(\hat{b}_0\text{cosh}(r)+\text{sinh}(r)\hat{a}_0^\dagger\right)\right]+\\
\text{sinh}(s)\left[\sqrt{\eta_{p1}}\left(\hat{b}_0e^{-i\phi}\text{sinh}(r)+e^{-i\phi}\text{cosh}(r)\hat{a}_0^\dagger\right)-i\sqrt{1-\eta_{p1}}\hat{c}_0^\dagger\right]\Bigg\}.
\label{bf}
\end{multline}

We then give the detector for the probe (conjugate) arm a classical amplification $A_p$ ($A_c$) and a homodyne phase $\phi_p$ ($\phi_c$). For the truncated SU(1,1) interferometer, we take $A_p=A_c=1$.

\subsubsection{Quadrature detection}
For homodyne detection, we define the quadrature operators $\hat{j}_p=A_p\left(e^{-i\phi_p}\hat{a}_f+e^{i\phi_p}\hat{a}_f^\dagger\right)$ and $\hat{j}_c=A_c\left(e^{-i\phi_c}\hat{b}_f+e^{i\phi_c}\hat{b}_f^\dagger\right)$ for the probe and conjugate arms, respectively. The joint quadrature operator is $\hat{J}=\hat{j}_p+\hat{j}_c$. The sensitivity $\Delta\phi$ is then calculated with the inputs and the operator transformations in Eqs.~\ref{af} and ~\ref{bf} from the variance (see Eq.~(1) in the main paper)
\begin{equation}
\left(\Delta^2\phi\right)_{J}=\frac{\left<J^2\right>-\left<J\right>^2}{\left|\partial_\phi\left<J\right>\right|^2}.
\label{quadsensivar}
\end{equation}
For $s=0$ \textit{and} $s=-r$, $A_p=A_c=1$, $\eta_{p1}=\eta_{c1}=\eta$, $\eta_{p2}=\eta_{c2}=1$, and $\phi_c=\pi/2$, one obtains Eq.~(3) in the main paper, which defines the sensitivity for both the full ($s=-r$) and truncated ($s=0$) SU(1,1) interferometers for these parameters. In the case $\eta=1$, Eq.~(\ref{quadsensivar}) gives rise to curves (i) and (v) in Fig.~2 of the main paper, where $r=\text{cosh}^{-1}(\sqrt{G})$ for gain $G$.

\subsubsection{Direct detection}
For direct intensity detection, we define the number operators $\hat{n}_p=A_p\hat{a}_f^\dagger\hat{a}_f$ and $\hat{n}_c=A_c\hat{b}_f^\dagger\hat{b}_f$ for the probe and conjugate arms, respectively, where we simply take $A_p$ and $A_c$ to be 1 or 0 depending on which detector(s) are on/off. We are interested in the sum $\hat{N}=\hat{n}_p+\hat{n}_c$. The phase sensitivity using direct detection is then calculated from the square root of the variance
\begin{equation}
\left(\Delta^2\phi\right)_{N}=\frac{\left<N^2\right>-\left<N\right>^2}{\left|\partial_\phi\left<N\right>\right|^2}.
\end{equation}
To derive the case of a single detector in the conjugate arm, we take $A_p=0$. In the case of no loss ($\eta_{p1}=\eta_{c1}=\eta_{p2}=\eta_{c2}=1$), the variance of the phase estimation using number detection for just the conjugate detector, with $s=-r$ and $A_c=1$ and at the best operating point, is
\begin{equation}
\left(\Delta^2\phi\right)_{N, \text{conj}}=\frac{\text{csch}^2(2r)}{|\alpha|^2},
\label{conjNsensi}
\end{equation}
which gives rise to curve (iii) in Fig.~2 of the main paper. Equation~(\ref{conjNsensi}) is also equivalent to $\left(\Delta^2\phi\right)_{J}$ with $A_p=0$ (only the conjugate homodyne detector) at the point of optimal sensitivity, corresponding to curve (ii) in Fig.~2 of the main paper.

For detecting both modes from the interferometer, we take $A_p=A_c=1$. In the case of no loss and $s=-r$, the variance of the phase estimation using number detection at the best operating point is
\begin{equation}
\left(\Delta^2\phi\right)_{N, \text{probe+conj}}=\frac{\text{csch}^4(2r)\left[2\text{cosh}(4r)+\sqrt{\text{cosh}(8r)}-1\right]}{2|\alpha|^2},
\end{equation}
which gives rise to curve (iv) in Fig.~2 of the main paper.

\subsection{Operating with the best sensitivity}
\begin{figure}
\centering
\includegraphics[scale=0.48]{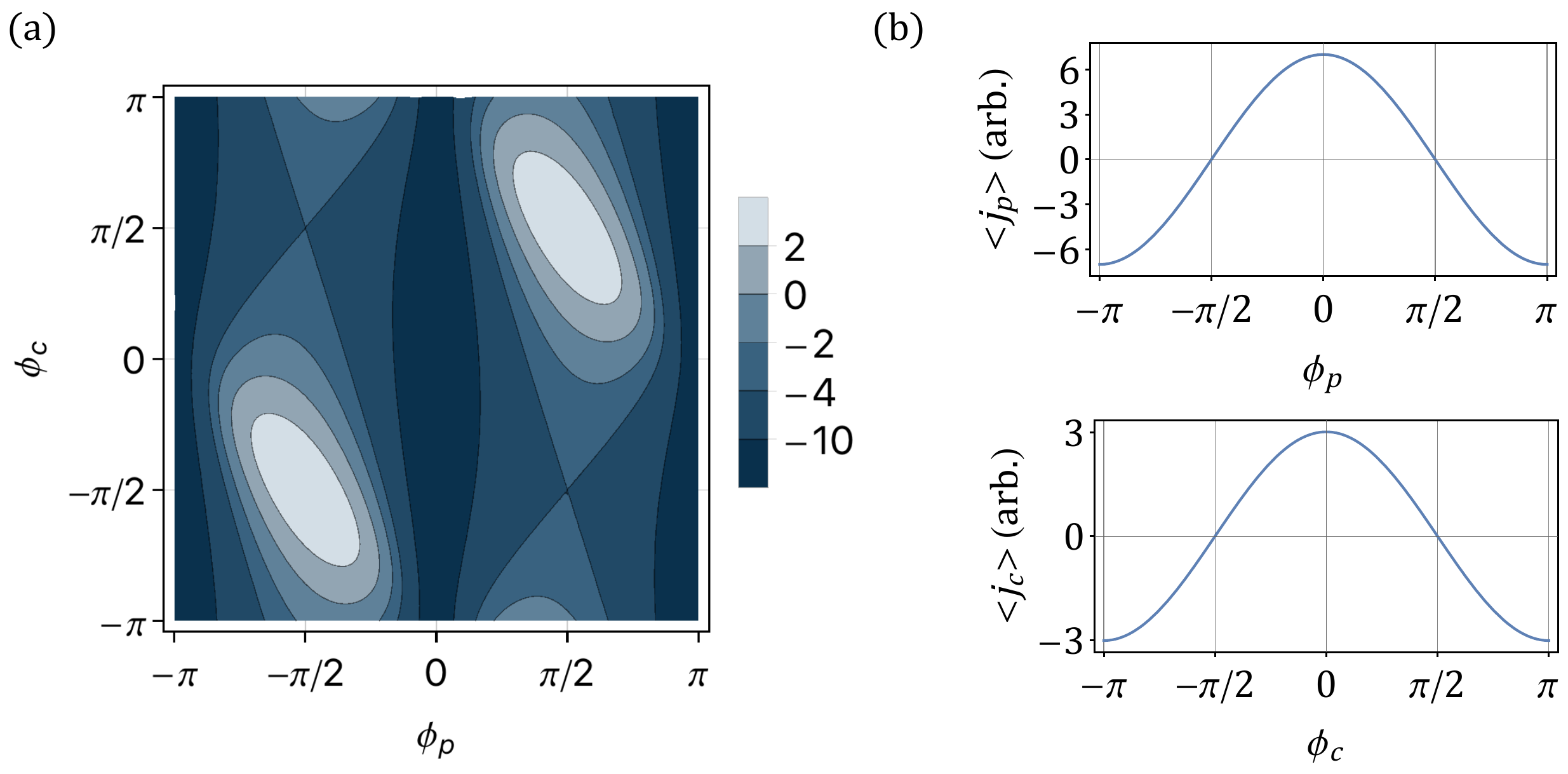}
\caption{(a) The SNRI (dB scale defined in legend) as a function of HD phases $\phi_p$ and $\phi_c$ in the case where $|\alpha|^2\gg1$ and $r=0.46$, which corresponds to 4~dB of squeezing. The regions of best phase sensitivity are shown in light colors. (b) The probe and conjugate HD signals ($\left<j_p\right>$ and $\left<j_c\right>$) as functions of the relative phases $\phi_p$ and $\phi_c$ considered in (a). This shows that the best SNRI is achieved by locking $\phi_p$ and $\phi_c$ to the zero crossing point with the same slope.}
\label{HDphaselockexplain}
\end{figure}

To analyze the signal-to-noise ratio improvement (SNRI) as a function of operating point, it is useful to determine the SNRI as a function of the relative homodyne phases $\phi_p$ and $\phi_c$. One can show from Eq.~\ref{quadsensivar} that
\begin{equation}
\Delta^2\phi_{\text{tSUI}}=\frac{2\eta+\left(1-2\eta\right)\text{sech}^2\left(r\right)+2\eta~\text{cos}\left(\phi_p+\phi_c\right)~\text{tanh}\left(r\right)}{2|\alpha|^2\eta~\text{sin}^2(\phi_p)}.
\label{snrifrac}
\end{equation}
In the special case where $\phi_c=\pi/2$, this reduces to Eq.~3 in the main paper. From Eq.~\ref{snrifrac} and the definitions of $\Delta^2\phi_{\text{coh}}$ and SNRI in the main paper, one obtains Fig.~\ref{HDphaselockexplain}(a), which shows the SNRI as a function of $\phi_p$ and $\phi_c$ for $\eta=1$. Figure~\ref{HDphaselockexplain}(b) shows the corresponding $\left<j_p\right>$ and $\left<j_c\right>$. Thus, the best SNRI occurs when $\phi_p$ and $\phi_c$ are locked to their phase quadratures (\textit{i.e.}, the zero-crossings of $\left<j_p\right>$ and $\left<j_c\right>$), and when $\left<j_p\right>$ and $\left<j_c\right>$ have the same slope. We investigate the SNRI as a function of operating point in the main paper by locking $\phi_c$ to its phase quadrature and scanning $\phi_p$.

\subsection{Discussion of losses}
As was shown in Refs.~$\left[\text{S2,~S3}\right]$, the sensitivity of the full SU(1,1) interferometer is inhibited by two types of losses: internal ($\eta_{p1}$, $\eta_{c1}$) and external ($\eta_{p2}$, $\eta_{c2}$). Examples of external losses include detector inefficiencies or imperfect visibility in homodyne detection, and internal losses include all losses inside and between the 4WM processes, such as absorption, optical loss, and imperfect mode-matching in the second 4WM process. One can show that internal and external losses have different effects on the sensitivity, and that internal losses are more detrimental~$\left[\text{S2}\right]$. With $\eta_{p1}=\eta_{c1}=\eta_{\text{int}}$ and $\eta_{p2}=\eta_{c2}=\eta_{\text{ext}}$, the variance of the phase estimation $\left(\Delta^2\phi\right)_{J}$ depends differently on $\eta_{\text{int}}$ vs.~$\eta_{\text{ext}}$. Specifically, for no external losses,
\begin{equation}
\left(\Delta^2\phi\right)_{J\text{, }\eta_{\text{ext}}=1}=\frac{e^{-r}\text{sech}\left(r\right)\left[1+\text{tanh}\left(r\right)-2\eta_{\text{int}}\text{tanh}\left(r\right)\right]}{2\eta_{\text{int}}|\alpha|^2}.
\end{equation}
For no internal losses,
\begin{equation}
\left(\Delta^2\phi\right)_{J\text{, }\eta_{\text{int}}=1}=\frac{2}{\eta_{\text{ext}}|\alpha|^2\left[1+\text{cosh}\left(2r\right)+\text{sinh}\left(2r\right)\right]^2}.
\end{equation}
For a gain of 3.3 ($r\approx1.15$), $|\alpha|^2\left(\Delta^2\phi\right)_{J\text{, }\eta_{\text{ext}}=1}$=0.05 for $\eta_{\text{int}}=0.8$, and $|\alpha|^2\left(\Delta^2\phi\right)_{J\text{, }\eta_{\text{int}}=1}$=0.02 for $\eta_{\text{ext}}=0.8$. Thus, the sensitivity is more robust against external losses than internal losses. The truncated SU(1,1) interferometer offers an advantage over the full version in that it eliminates any internal loss associated with the second 4WM process. A disadvantage of the truncated SU(1,1) interferometer compared to the full version is that all losses are internal, so that one must use high quantum efficiency detectors and achieve high homodyne visibilities to minimize loss.

\subsection{Verifying the SQL}
We perform auxiliary experiments to compare the SNR measured with coherent beams to that expected for the truncated Mach-Zehnder interferometer. For our two-homodyne setup, we expect $\text{SNR}_{\text{coh}}=(\delta\phi)^2N_p$, where $\delta\phi$ is the RMS amplitude of the EOM phase modulation and $N_p=2\eta_{\text{coh}}\rho P/eB$, where $\eta_{\text{coh}}$ is a loss parameter, $\rho=0.64~\text{A/W}$ is the responsivity for an ideal (100~\% quantum efficiency) detector at 795~nm, $P=400\pm20~\text{nW}$ is the power in the probe beam, $e$ is the electric charge, and $B$ is the ``equivalent noise bandwidth.'' The loss parameter $\eta_{\text{coh}}$ includes the detector quantum efficiency 0.9, the homodyne visibility ($\approx0.95$ for these experiments), and separation from electronic noise ($20~\text{dB}$ separation, hence negligible), from which we expect $\eta_{\text{coh}}\approx0.8$. This $\eta_{\text{coh}}$ ($\approx20~\%$ loss) differs from the loss parameter $\eta=0.65$ in the main paper ($\approx35~\%$ loss) because it only accounts for detector loss, whereas $\eta$ also includes losses in the quantum state preparation. The apparent SNR shown on the spectrum analyzer trace (Fig.~4a in the main paper) needs a correction to account for the different ways in which broadband and narrowband signals are processed (see for example Ref.~$\left[\text{S4}\right]$) and the resolution bandwidth converted to bandwidth to account for the filter function employed. We use the built-in software corrections on the spectrum analyzer to determine the corrected SNR in a bandwidth of 30~kHz.

In the case where $\delta\phi=1.7\pm0.2~\text{mrad}$, we find that our corrected $\text{SNR}_{\text{coh}}\approx22.5~\text{dB}$. Given the uncertainties in $\delta\phi$, $P$, and the spectrum analyzer calibration, this agrees with the expected $\eta_{\text{coh}}\approx0.8$. Thus we conclude that the measured SQL agrees with the expected limit for a truncated Mach-Zehnder interferometer.

In the data shown in Fig.~4 of the main paper, the homodyne visibility was $\approx98~\%$, and the electronic noise floor separation was $\approx18~\text{dB}$, which corresponds to detector losses of $\approx~15\%$. If the coherent state measurements could have been made with a perfect detector, then the $\text{SNR}_{\text{coh}}$ would have increased by $\approx1~\text{dB}$, and the corresponding SNRI over the idealized $\text{SNR}_{\text{coh}}$ would still have been $3~\text{dB}$.

\vspace{4mm}
\noindent{[S1]~Brian E. Anderson et al., to be published.}\\
\noindent{[S2] A. M. Marino, N. V. Corzo Trejo, and P. D. Lett, Phys. Rev. A 86, 023844 (2012).}\\
\noindent{[S3] F. Hudelist, J. Kong, C. Liu, J. Jing, Z. Y. Ou, and W. Zhang, Nat. Commun. 5, 3049 (2014).}\\
\noindent{[S4] Agilent Technologies, Inc. ÒAgilent Spectrum and Signal Analyzer Measurements and Noise, Application Note.Ó 5966-4008E (2012).}\\

\end{document}